\documentclass[10pt,letterpaper]{article}
\usepackage[top=0.85in,left=2.75in,footskip=0.75in,marginparwidth=2in]{geometry}

\usepackage[utf8]{inputenc}
\usepackage{cite}
\usepackage{nameref,hyperref}
\usepackage{amssymb}
\usepackage{graphicx}
\usepackage{epstopdf}
\usepackage{microtype}
\DisableLigatures[f]{encoding = *, family = * }

\raggedright
\usepackage{changepage}

\usepackage[aboveskip=1pt,labelfont=bf,labelsep=period,singlelinecheck=off]{caption}

\makeatletter
\renewcommand{\@biblabel}[1]{\quad#1.}
\makeatother

\usepackage{lastpage,fancyhdr,graphicx}

\fancyheadoffset[L]{2.25in}
\fancyfootoffset[L]{2.25in}

\usepackage{graphicx}
\usepackage[pscoord]{eso-pic}
\usepackage[fulladjust]{marginnote}
\reversemarginpar

\begin{document}
\vspace*{0.35in}

\begin{flushleft}
{\Large
\textbf\newline{Magnetocaloric effect in the high-temperature antiferromagnet YbCoC$_2$}
}
\newline
\\
D. A. Salamatin\textsuperscript{1,2,*},
V. N. Krasnorussky\textsuperscript{1},
A. V. Semeno\textsuperscript{1,3},
A. V. Bokov\textsuperscript{1},
A. Velichkov\textsuperscript{2,4},
Z. Surowiec\textsuperscript{2,5},
A. V. Tsvyashchenko\textsuperscript{1}
\\
\bigskip
\bf{1} Vereshchagin Institute for High Pressure Physics, Russian Academy of Sciences, Kaluzhskoe shosse 14, Troitsk, Moscow, Russia
\\
\bf{2} Joint Institute for Nuclear Research, 6 Joliot-Curie Str., Dubna, Russia
\\
\bf{3} Prokhorov General Physics Institute, Russian Academy of Sciences, 38 Vavilov Str., Moscow, Russia
\\
\bf{4} Institute for Nuclear Research and Nuclear Energy, 72 Tsarigradsko shosse Blvd., Sofia, Bulgaria
\\
\bf{5} Institute of Physics, M. Curie-Sklodowska University, pl. M. Curie-Sklodowskiej, 1, Lublin, Poland
\\
\bigskip
* dasalam@gmail.com

\end{flushleft}

\section*{Abstract}
The magnetic $H$-$T$ phase diagram and magnetocaloric effect in the recently discovered 
high-temperature heavy-fermion compound YbCoC$_2$ have been studied. With the increase in the external 
magnetic field YbCoC$_2$ experiences the metamagnetic transition and then transition to the
ferromagnetic state. The dependencies of magnetic entropy change -$\Delta S_m (T)$ have segments 
with positive and negative magnetocaloric effects for $\Delta H \leq 6$~T. For 
$\Delta H = 9$~T magnetocaloric effect becomes positive with a maximum value of -$\Delta S_m (T)$ is 
4.1 J / kg K and a refrigerant capacity is 56.6 J / kg.

\section{Introduciton}
\label{sec:intro}
Recently, the compounds of GdCoC$_2$, GdNiC$_2$, NdRhC$_2$ and PrRhC$_2$ 
have been predicted to be topological Weyl semimetals (TWS) \cite{RNiC2_TWS_npj2022}.
In these compounds, inversion symmetry and time-reversal symmetry 
are broken due to the noncentrosymmetric orthorhombic structure of the CeNiC$_2$-type
and low-temperature magnetic transitions, respectively. 
The unique properties of magnetic TWS can be extremely useful
in the context of information technology (e.g., quantum computing), 
given that such massless charged particles
will carry electric current without Joule heating \cite{Spin-transfer_TWS_SciRep2019}.
The theory provides a clear guide to the implementation of magnetic TWS, 
but so far, there are only a few experimentally confirmed examples
of TWS with time-reversal symmetry breaking \cite{EuCd2As2_mTWS_ScienceAdv2019, YbMnBi2_mTWS_Nature2019}.

The $R$CoC$_2$ compounds with magnetic $R$ have ferromagnetic (FM)
order at low temperatures \cite{Schafer1990, Amanai1995, HoCoC2_HoNiC2_JMMM2017, RCoC2_JALCOM1996}.
It is believed that in these compounds Co ions do not have a magnetic moment 
and interactions beyond Ruderman-Kittel-Kasuya-Yosida interaction
make a significant contribution to the stabilization of magnetism \cite{Schafer1997}.

One of the promising applications of rare-earth-based magnets is their use 
as cooling refrigerators. This becomes possible due to 
their large magnetocaloric effect (MCE).
The MCE effect is the result of a change in the entropy of 
magnetic spins $\Delta S_m$ under the influence of a magnetic field
and can be fully characterized by a change in temperature 
in an adiabatic process ($\Delta T_m$)
and a change in magnetic entropy in an isothermal process ($\Delta S_m$) 
depending on a change in an external magnetic field.
$\Delta S_m$ can be obtained indirectly from isothermal measurements 
of the magnetization.
Previously, large reversible MCE in HoCoC$_2$, 
ErCoC$_2$ \cite{HoCoC2_ErCoC2_MCE_Intermetallics2017},
TbCoC$_2$ \cite{TbCoC2_MCE_APL2008} and giant reversible 
MCE in TWS GdCoC$_2$ \cite{GdCoC2_MCE_RSC2016} have been observed 
using such measurements.

The moderately heavy fermion compound YbCoC$_2$ ($\gamma$ = 190 mJ/mol-K$^2$) 
has an antiferromagnetic (AFM) transition at $T_{\mathrm{N}}$ = 27 K, 
which is the highest temperature for the Yb-based magnetic compounds.
The magnetic structure of YbCoC$_2$ in a zero magnetic field at 
$T$ = 1.3 -- 27 K is a sine-modulated AFM structure with wave vector (0, 0, $k_z$),
where $k_z$ depends on temperature: $k_z = 0.28$ at $T_{\mathrm{N}}$ 
and locks-in to the value of 1/4 below 8 K.
The magnetic moment amplitude of Yb ions is $\mu_{\mathrm{Yb}} = 1.32$ $\mu_B$ 
at 1.3 K, which is lower than the full magnetic moment of a Yb$^{3+}$ free ion 
($gJ = 4$ $\mu_B$) \cite{YbCoC2_PRB2020}.

\section{Methods}
\label{sec:method}
A polycrystalline single-phase sample of YbCoC$_2$ was synthesized using
high pressure-high temperature technique
at $P = $ 8 GPa and $T = $ 1500 -- 1700 K using
Toroid high-pressure cell 
by melting Yb, Co, C and characterized in Ref.~\cite{YbCoC2_PRB2020}.

Magnetic moment measurements were made on the VSM option of PPMS, Quantum Design.
The isothermal magnetization curves were obtained by 
increasing the magnetic field from 0 to 9 T and changing the temperature
from 2 to 80 K under field cooling conditions with variable temperature steps: $\delta T$ = 4 K 
above and well below $T_{\mathrm{N}}$ and $\delta T$ = 2 K near $T_{\mathrm{N}}$.
The magnetic field step was held at $\mu_0 \delta H$ = 0.1 T.

\section{Results and discussion}
\label{sec:method}

\begin{figure}[h]
\centering
\includegraphics[width=1.0\columnwidth]{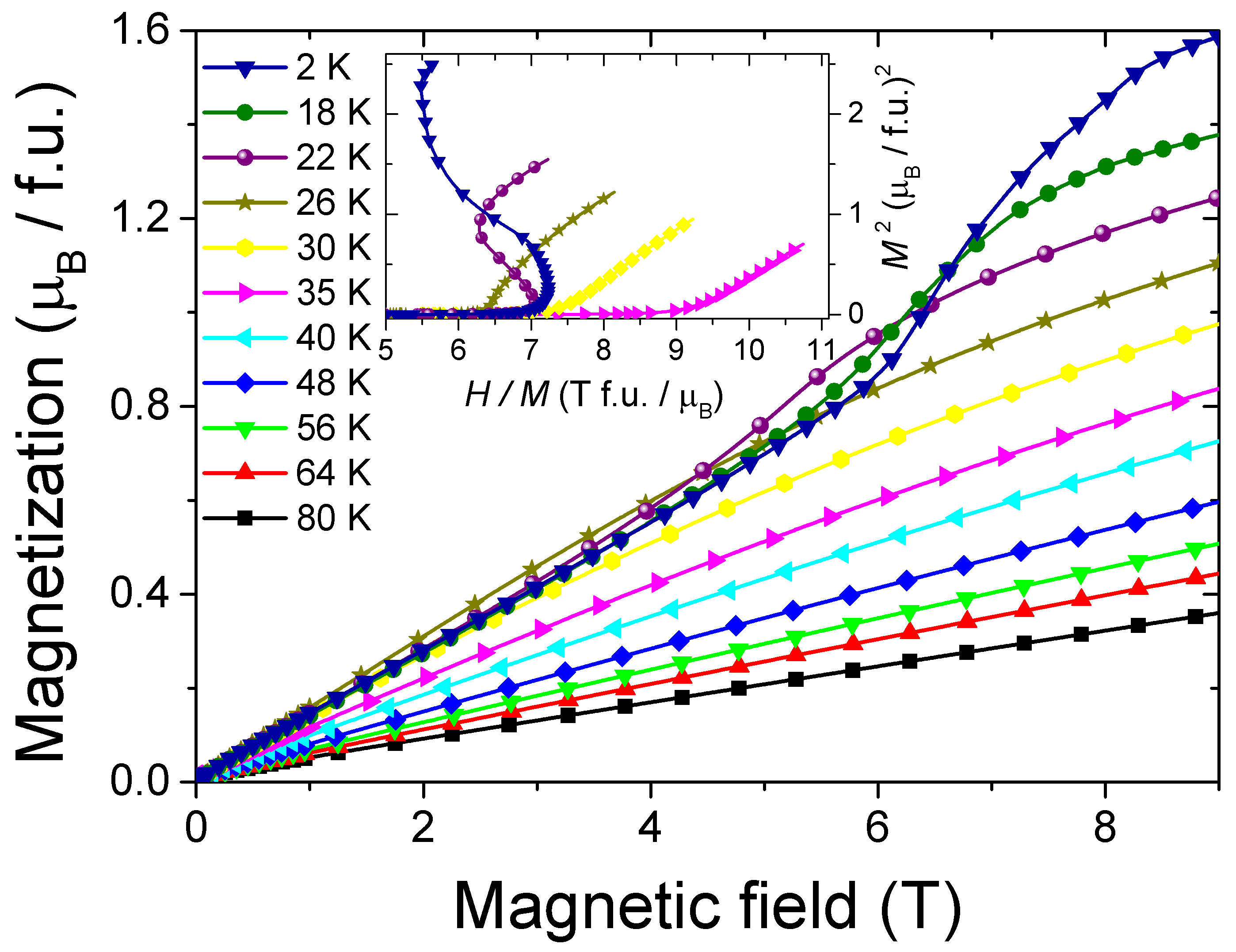}
\caption{Isothermal magnetization dependencies $M(H)$ of YbCoC$_2$.
Inset: Arrot plots ($M^2$ vs $H / M$) for selected temperatures.}
\label{fig1}
\end{figure}

Some $M(H)$ dependencies from the temperature range 2 -- 80 K are shown on Fig. \ref{fig1}.
A detailed analysis of dependencies at $T < T_{\mathrm{N}}$ and 
in low fields shows that the $M(H)$ curves have almost linear slopes
and the magnetization increases with increasing temperature. 
Also, spontaneous magnetization is absent at all temperatures.
This behavior is characteristic of an AFM material. 
At high magnetic fields ($H \geq H_{c1}$), there is a sharp 
increase in $M(H)$ associated with a metamagnetic transition.
For this transition, hysteresis is observed in the $M(H)$ 
curves \cite{YbCoC2_PRB2020} (not shown here).
The metamagnetic transition is not observed in the paramagnetic region 
($T > T_{\mathrm{N}}$). Therefore, its nature 
is connected with the AFM structure of Yb magnetic moments.

$M(T)$ measured in magnetic field of 7 T (see Fig. \ref{fig2}) 
demonstrates a plateau-like behavior at $T \lesssim 24$ K
(this temperature is defined as a minimum in $dM/dT$ vs $T$ dependence), 
which corresponds to the transition from the paramagnetic (PM) 
to the ferromagnetic state.
We have plotted the magnetic $H$-$T$ phase diagram of YbCoC$_2$ 
obtained from the magnetization measurements (see the inset of Fig. \ref{fig2}).
$H_{c1}$ and $H_{c2}$ were determined as local maxima in $dM / dH$ vs $T$ 
and from $M(T)$ dependencies.
$H_{c1}$ is associated with the metamagnetic transition to the 
intermediate magnetic phase (IM), and $H_{c2}$ 
is probably associated with the transition of
YbCoC$_2$ to the FM state induced by the external magnetic field 
in which the magnetic unit cell has a finite value of magnetization.

Some Arrot curves have a negative slope (see the inset of Fig. \ref{fig1}).
According to the Banerjee criterion \cite{PhaseTrans_Banerjee1964}, 
this indicates that the metamagnetic transition is of first-order type.

The saturation of magnetization is not observed in fields of 9 T 
at all temperatures. $M$(9 T) $\approx$ 35 J / T-kg at $T$ = 2 K, 
which corresponds to a magnetic moment of about 1.6 $\mu_{\mathrm{B}}$ / f.u.
This value is smaller than the saturation magnetic moment 
of Yb$^{3+}$ ion ($m_s = 4.0 \mu_{\mathrm{B}}$).

\begin{figure}[h]
\centering
\includegraphics[width=1.0\columnwidth]{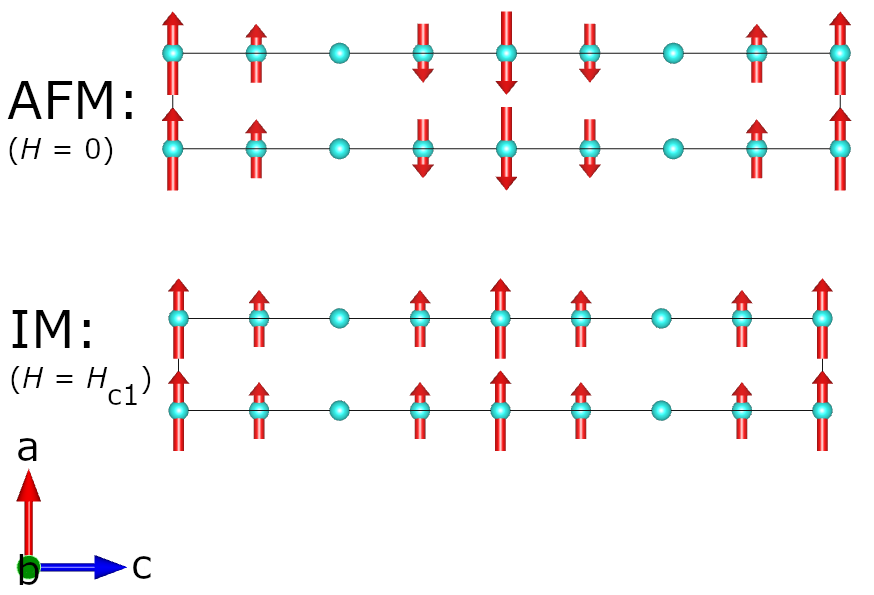}
\caption{Possible transformation of YbCoC$_2$ magnetic structure 
in the external magnetic field: from the antiferromagnetic sine-wave (AFM) at zero field to
the intermediate spin-polarised (IM) at $H = H_{c1}$. Only the Yb ions (blue circles) and their 
magnetic moments (red arrows) are shown.}
\label{AFM-to-IM}
\end{figure}

\begin{figure}[h]
\centering
\includegraphics[width=1.0\columnwidth]{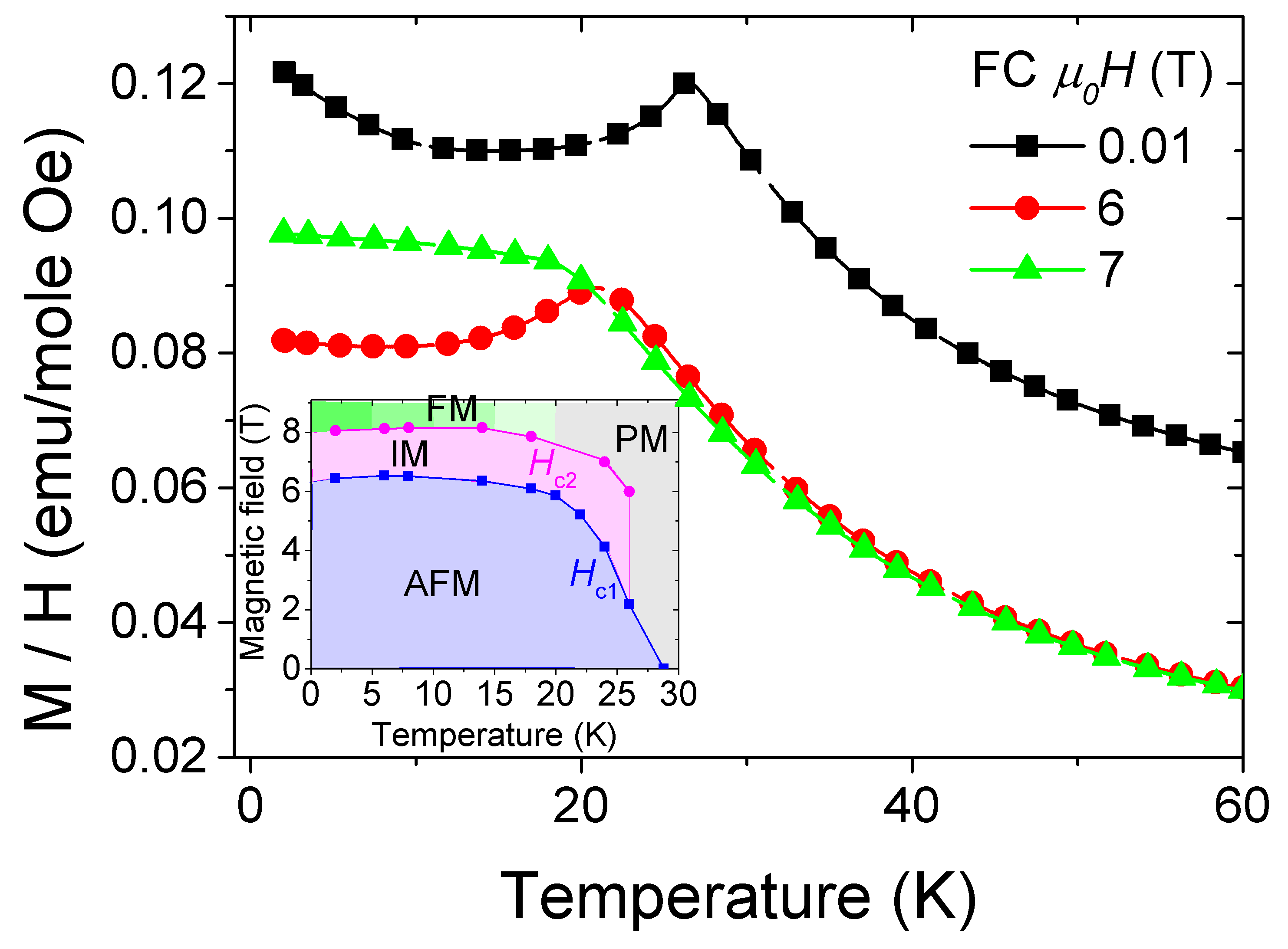}
\caption{
  Temperature dependencies of $M / H$ measured in various external magnetic fields in
  field-cooled mode.
  Inset: possible magnetic $H$-$T$ phase diagram of YbCoC$_2$ 
  (where AFM - antiferromagnetic phase (blue area),
  IM - intermediate magnetic phase corresponding to the metamagnetic phase transition (pink area) 
  and FM - ferromagnetic phase (green area)).}
\label{fig2}
\end{figure}

It is interesting to note that the mean value of the
magnetic moment of the halve positive period of Yb moments sine-wave 
- the magnetic structure of YbCoC$_2$ at $T = 2$~K and in zero magnetic field (see Fig. \ref{AFM-to-IM}),
is $1.32 \cdot (1 + 2 \cdot sin(\pi/4))$ / 4 = 0.8 $\mu_B$ / f.u.
$M(H)$ at $T = 2$~K reaches this value at $H_{c1}$. 
Hence it can be assumed that there is a smooth, almost linear, transformation 
from the sine-wave modulation to the spin-polarised magnetic structure for $H$ = 0 -- $H_{c1}$ 
with preservation of the Yb mean magnetic moment (see Fig. \ref{AFM-to-IM}). 

As seen from Fig. \ref{fig1} the $M(H)$ exceeds the value of 0.8 $\mu_B$ / f.u for $H > H_{c1}$ 
in the magnetically ordered state. So the mean magnetic moment increases in value above $H_{c1}$,
which may be connected with the increase of Yb mean magnetic moment and the additional contribution from Co.

The temperature dependence of the magnetic susceptibility behaves anomalously at $H$ = 100 Oe (not shown). 
$\chi^{-1}(T)$ decreases slightly above 350 K, similar to the GdCoC$_2$ compound \cite{GdCoC2_Matsuo1996}. 
It may be connected with the finite value of Co magnetic moment in GdCoC$_2$ and YbCoC$_2$.
The finite value of Co magnetic moments in YbCoC$_2$ was predicted 
by numerical calculations \cite{YbCoC2_PRB2020}.

$\Delta S_m(T)$ and its relative error for different $\Delta H = H_f - H_i$
were determined from the isothermal magnetizations using well-known numerical methods \cite{MCE_Pecharsky_JAP1999}.

\begin{equation}
  \Delta S_m(T) = \int_{H_i}^{H_f} \left( \frac{\partial M}{\partial T} \right)_H dH ,
\end{equation}
where $H_i$ (= 0 in the present case), 
$H_f$ are the initial and final values of the applied magnetic fields,
respectively. The relative error of $\Delta S_m(T)$ did not exceed 10 \%.

\begin{figure}[h]
\centering
\includegraphics[width=1.0\columnwidth]{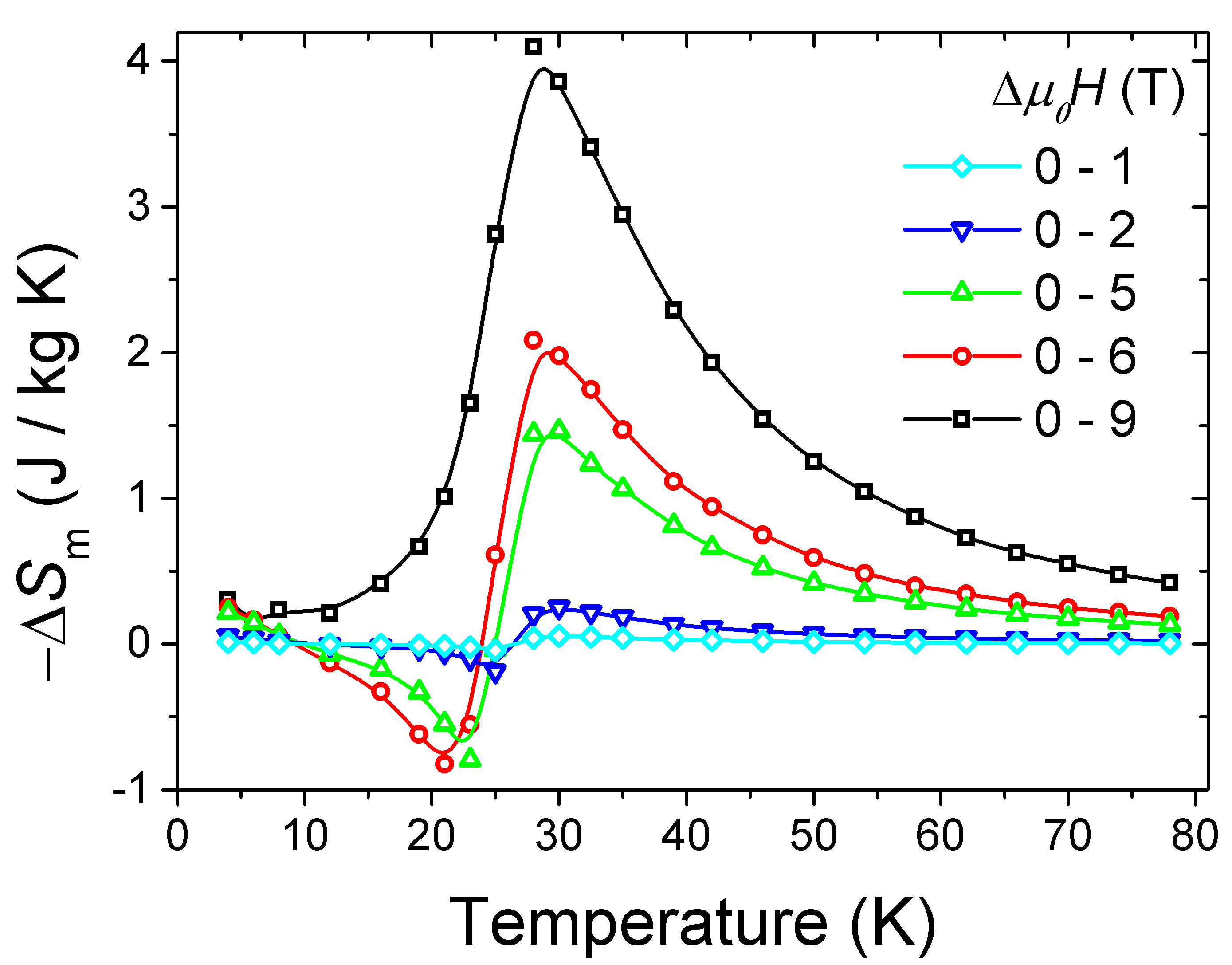}
\caption{
  Temperature dependencies of the magnetic entropy change $-\Delta S_m(T)$ 
  for YbCoC$_2$ for different magnetic field changes
  (points - experimental data, lines - spline fits).
}
\label{fig3}
\end{figure}

The dependencies $-\Delta S_m (T)$ for $\Delta H \leq$ 0 -- 6 T 
have segments with positive and negative MCE (see Fig. \ref{fig3}).
These dependencies have minima, which are connected with 
the AFM nature of the magnetic ordering in YbCoC$_2$ in low magnetic fields.
The $-\Delta S_m$ minima shift towards lower temperatures 
with an increase of $\Delta H$ in accordance with the phase diagram 
(the inset of Fig. \ref{fig2}) and the positive MCE appears at $T < 10$ K 
for $\Delta H \geq$ 0 -- 5 T.
For $\Delta H =$ 0 -- 9 T a strong external magnetic field 
suppresses the AFM structure, and MCE becomes positive in the full temperature range
and $-\Delta S_m(T)$ has a typical FM 
"caret-like" shape \cite{MCE_RevMatSci2000}.
$-\Delta S_m(T)$ reaches a maximum of 4.1 J / kg-K at $T \approx 28$ K. 

The amount of heat that can be transferred between the cold 
and hot parts in one cooling cycle is 
RC = $\int_{T_1}^{T_2} |\Delta S_m| dT$ = 56.6 J / kg
for $\Delta H = $ 0 -- 9 T. Here, $T_1$, $T_2$ are 
the temperatures corresponding to both sides of 
the half maximum value of $-\Delta S_m(T)$.

If we compare the obtained $-\Delta S_m$ maxima for two TWS, 
YbCoC$_2$ and GdCoC$_2$, we get max$(- \Delta S_m)$ / $R ln(2J + 1)$ = 0.18 
and 0.46, respectively.

\section{Conclusion}
\label{sec:conclusion}

In conclusion, we have plotted the magnetic $H$-$T$ phase diagram of YbCoC$_2$
in the magnetic field range 0--9~T and the temperature range 2--30~K. 
It is shown that with an increase in the external magnetic field, 
the metamagnetic transition of YbCoC$_2$ to the IM phase occurs, 
and with a further increase of magnetic field, it goes to the FM phase. 
The magnetic structure of the IM and FM phases requires further research 
by means of neutron diffraction in the external magnetic field and
magnetization measurements on single crystal.
The smooth transformation from the sine-wave modulation to the spin-polarised 
magnetic structure for $H$ = 0 -- $H_{c1}$ was observed. While further increases in magnetization
in higher magnetic field is connected with the increase of 
Yb mean magnetic moment, and the additional contribution from Co magnetic moments.
The MCE for YbCoC$_2$ has been calculated for $\Delta H$ up to 9~T. 
Due to the AFM - FM transition, MCE in YbCoC$_2$ changes sign with the increasing of $\Delta H$.

\section*{Acknowledgements}
The authors are grateful to V. V. Brazhkin for support of the work and 
M. A. Anisimov for fruitful discussions.
This experimental research was funded by the Russian Science Foundation Grant No. 22-12-00008.
We are grateful for the support in performing magnetization measurements 
provided by the Polish representative at the Joint Institute for Nuclear Research.

\bibliography{main}

\bibliographystyle{abbrv}

\end{document}